\documentclass{sfm}

\usepackage[T2A]{fontenc}
\usepackage[english]{babel}
\usepackage{lscape}
\usepackage{graphicx}
\usepackage{amssymb}
\usepackage{bibspacing}

\usepackage{epsfig}

\def\figurename{Figure}
\def\tablename{Table}
\newcommand{\titl}[3]
{

\begin{flushleft}
\baselineskip 14pt
{\sffamily\bfseries\Large #1\\[2.1mm]

}
\baselineskip 12pt
{\large #2\\[2.1mm]

}
{\baselineskip 11pt
\small\it #3\\[5mm]

}
\end{flushleft}

}

\newcommand{\abstre}[1]{
{

\small\noindent
{\bfseries Abstract} \hspace{5mm}
\baselineskip 10pt
\noindent
#1

}}

\usepackage{graphicx}        % standard LaTeX graphics tool, added byTL


\begin{document}

\def\llm{{\sc LLmodels}}
\def\atl{{\sc ATLAS9}}
\def\aatl{{\sc ATLAS12}}
\def\starsp{{\sc STARSP}}
\def\aur{$\Theta$~Aur}
\def\logg{\log g}
\def\tauros{\tau_{\rm Ross}}
\def\kms{km\,s$^{-1}$}
\def\bz{$\langle B_{\rm z} \rangle$}
\def\degr{^\circ}
% journals
\def\aaps{A\&AS}
\def\aap{A\&A}
\def\apjs{ApJS}
\def\apj{ApJ}
\def\rmxaa{Rev. Mexicana Astron. Astrofis.}
\def\mnras{MNRAS}
\def\actaa{Acta Astron.}
\newcommand{\Tef}{T$_{\rm eff}$~}
\newcommand{\Vt}{$V_t$}
\newcommand{\CC}{$^{12}$C/$^{13}$C~}
\newcommand{\CDC}{$^{12}$C/$^{13}$C~}

%%%%%%
\pagebreak

\thispagestyle{titlehead}

\setcounter{section}{0}
\setcounter{figure}{0}
\setcounter{table}{0}

%%%%%%%%%%%%%%%%%%%%%%%%%%%

\markboth{Lueftinger et al.}{Doppler- and Magnetic Doppler imaging of Ap stars}

\titl{Simultaneous mapping of chemical abundances and magnetic field structure in Ap stars}{Lueftinger T.$^1$}
{$^1$University of Vienna, Department of Astrophysics, Austria, \\email: {\tt theresa.rank-lueftinger@univie.ac.at} 
 }

\abstre{
Magnetic A stars represent about 5\,\% of 
the upper main sequence stars and exhibit highly ordered, 
very stable and often very strong magnetic fields. They frequently 
show both, brightness- and spectral line profile variations 
synchronised to stellar rotation, which are believed to be 
produced by atomic diffusion operating in the stars' atmospheres, 
that are stabilized by multi-kG magnetic fields.
In recent years, with the development and the application 
of the Doppler- and magnetic Doppler imaging 
technique and the availability of high precision spectroscopic 
and spectropolarimetric data, it has became possible to map 
chemical abundances and magnetic field structures of Ap stars simultaneously and 
in more and more detail, based on full Stokes vector observations. 
Here I will review the state-of-the-art in understanding 
Ap star spots and their relation to magnetic fields, 
the development of Doppler- and magnetic Doppler imaging into 
one of the most powerful astrophysical remote sensing methods 
and the physics of Ap stars atmospheres we can deduce from the 
simultaneous mapping of magnetic field structure and chemical abundances.
}

\baselineskip 12pt

\section{Introduction}
Chemically peculiar stars of the upper main sequence cover the whole range of spectral types from early B to early F and 
and do not seem to be constrained to a particular evolutionary stage (Kochukhov \& Bagnulo, 2006). 
In addition to their strong, globally strucured magnetic fields, many Ap stars also show prominent spectral line profile 
variations synchronized to stellar 
rotation, which, within the framework of the oblique rotator model (introduced
by Stibbs (1950), is attributed to oblique magnetic and rotation axes 
and to the presence of a non-uniform distribution of
chemical elements on their surface. With few exceptions such inhomogeneities exist only in the
atmospheres of A stars with magnetic fields, demonstrating that these fields play a
crucial role in their formation and evolution. 
The mentioned chemical peculiarities are attributed to the selective diffusion
of ions under the competitive action of radiative acceleration and gravitational
settling within an atmosphere that is stabilized 
by the magnetic field (Michaud, 1970). The observed field structures and the chemical spots typically remain stable 
over many decades. 

Modeling of magnetic field effects is extremely challenging, 
both from the observational and the theoretical point of 
view, and only recent groundbreaking advances in observational 
instrumentation, as well as a deeper theoretical understanding of 
magnetohydrodynamic processes and diffusion in stellar atmospheres (e.g. Alecian et al. 2011) enable us to model stellar fields 
and their relation to chemical inhomogeneities in more and more detail. 

At the same time, the development and application of magnetic Doppler imaging (Kochukhov et al. 2004, 
Donati et al. 2006, L\"uftinger et al. 2010a,b), a method that has become one of the most powerful remote sensing
methods in present astrophysics, allows a detailed confrontation 
of these new models with observations. 

\section{Doppler- and magnetic Doppler Imaging}

\subsection{Historic overview}
The first solution for determining surface anomalies from observed 
spectra was presented by Deutsch in 1958: they developed equivalent widths and magnetic potential 
into spherical harmonics and related Laplace coefficients of these expansions to 
Fourier coefficients of the observed curves. This approach was ingenious, but limited, as informations 
of line profile variations could not be used and surface resolution was poor. 
Pyper (1969), Rice (1970), and Falk \& Wehlau (1974) later on changed from using line strength 
variations to working with variations of the line profile shape, which contains much more 
information, and Doppler Imaging (DI) as we know it today, can be ascribed to this work.
During the mid 1970's, Russian astronomers (Goncharsky et al., 1977), for the first 
time solved the inverse problem, applying mathematical equations relating inhomogeneities 
of temperature or abundance on a stellar surface to time series of observed line profiles  
using the Tikhonov regularization method (Tikhonov, 1963). 
In the following years, the application of the according computer codes and 
a remarkable increase in data quality notably increased the potential of 
the DI technique. 
A collaboration between Rice, Wehlau and Khokhlova resulted in the mapping of several Ap stars (first 
results published for $\epsilon\,UMa$ in 1981 (Rice, Wehlau, Khokhlova \& Piskunov, 1981).  
Vogt, Penrod and Hatzes extended the Doppler imaging work to cool 
stars and presented a new inversion technique (Vogt, Penrod and Hatzes, 1987) based 
on the Maximum Entropy Regularization Method (MEMSYS).
In the following years, groups around Piskunov, Rice, and Wehlau, in collaboration 
with Tuominen and Strassmeier, and around Vogt, Penrod, and Hatzes refined DI programs and 
extended their application. Astronomers like Cameron (1990), Brown et al. (1991), or K\"urster 
(1993) contributed with significant Doppler Imaging work or published new computer codes. 
An important step was taken in the beginning of the 1990's: Magnetic (or Zeeman) Doppler Imaging (MDI, ZDI), 
is introduced, involving Stokes parameters I, Q, U, and V in the analysis, by Semel and Donati, 
in collaboration with Brown and Rees (Brown et al. 1991) and independently by
Piskunov and Rice in 1993. 

\subsection{Principles and techniques}

Within DI and MDI, time-series of high-resolution observations of stellar spectra, which are Doppler broadened 
and modulated due to stellar rotation are inverted into two-dimensional surface maps of parameters like elemental abundance 
(or temperature if applied to spectra of cool stars) and magnetic field geometry.  
From the mathematical point of view, during the inversion process, a total discrepancy 
function $\Psi = D + R$ is minimized, whereby $D$ characterizes the discrepancy between the 
observed and theoretical phase-resolved spectra, and $R$ is the regularization functional. 
This regularization functional ensures stability of the complex optimization algorithm within 
(magnetic) DI and the simplest possible and unique solution independent from the initial guess 
and the surface discretization. 
Various different codes, applied to cool and/or hot stars have been developed meanwhile. In the following we present 
a short overview of existing software and its authors and applications to early- and/or late type stars: 
\\\\
{\bf Late type stars:} 
\begin{itemize}
\item{{\sc DOTS}, Collier Cameron (1995, 1997): brightness, modified version extended towards MDI (=ZDI, designation mainly used for late tye
stars)}  
\item{{\sc TEMPMAP}, Rice \& Strassmeier (2000): temperature} 
\item{i{\sc MAP}, Kopf, Caroll et al. (2009): temperature}
\item{magnetic-imaging codes of Brown et al. (1991), Donati et al. (2001, 2006), Hussain et al. (2010), Petit et al. (2004):   
surface field, brightness, accretion powered excess} 
\item{{\sc INVERS13}: Kochukhov \& Piskunov (2009), Rosen \& Kochukhov, (2012): temperature and magnetic field simultaneously} 
\end{itemize}

{\bf Early type stars:}
\begin{itemize}
\item{{\sc INVERS7, INVERS8, INVERS10, INVERS12, INVERS13}, Piskunov \& Kochukhov (2002), Kochukhov \& Piskunov (2002), surface field,
abundances, simultaneously}
\item{magnetic-imaging code of Brown et al. (1991), Donati et al. (2001, 2006), Petit et al. (2011): surface field, abundances, 
separately}
\end{itemize}

\subsection{Abundances and magnetic field structure in Ap stars}
In recent years, there has been huge progress in the number and level of detail in which the surfaces of Ap stars have been 
mapped (e.g. by Hatzes et al., Kochukhov et al., Lueftinger et al., Donati  et al., Petit et al., Rice  et al., Silvester et al., 
Wehlau et al., etc., just to name a few examples). 

In particular with the advent of new generation spectropolarimters like, e.g., ESPaDOnS, NARVAL, and HARPSpol, the considerable
increase in computing power and the use of efficient numerical algorithms it is now possible to take full advantage of
inversions based on data sets in all Stokes parameters. As has been shown for, e.g., $\alpha^2$\,CVn (Kochukhov \& Wade
2010), these datasets have the potential to unveal far more complex surface magnetic field components than evident from only 
Stokes I and V inversions (see Fig.\,2). 

The above mentioned studies, surprisingly also reveal a pronounced diversity of spot structures, even for elements found within 
comparable positions in the periodic table of elements. The long considered 'typical' Ap star spot pattern of iron peak 
elements distributed around the magnetic equator and rare-earth elements (REE) primarily found at the poles of a field dominated 
by the dipolar component is not confirmed. Actually only few elements like, e.g., Li, O, or Eu seem to exhibit a consistent 
correlation to the dipolar field in terms of a well defined spot or ring structure - most other elements do not (obviously) 
show such a correlation. Potentially these elements are more sensitive to the complex horizontal magnetic field component as
revealed in Stokes IQUV studies of 53 Cam (Fig.\,1) and $\alpha^2$\,CVn (Fig. 2, Kochukhov \& Wade 2010, Kochukhov et al. 2004). Both stars exhibit an
overall field topology, that is roughly dipolar and dominated by the radial field component. The field intensity distribution,
though, shows a much higher level of complexity, which is cauesd by small-scale variations of the horizontal magnetic field.  
Both analysis, taking advantage of a full set of Stokes IQUV observations unveil a far more complex magnetic field intensitiy distribution than originally expected for Ap stars, but 
they still show an intrinsically different surface magnetic field complexity. 
It is still not clear, where this difference arises from, and if/how it is connected to basic stellar properties. 
Similar studies of a significantly larger sample of stars are indispensable for a better understanding.

\begin{figure}[!t]
\begin{center}
 \includegraphics[width=8cm]{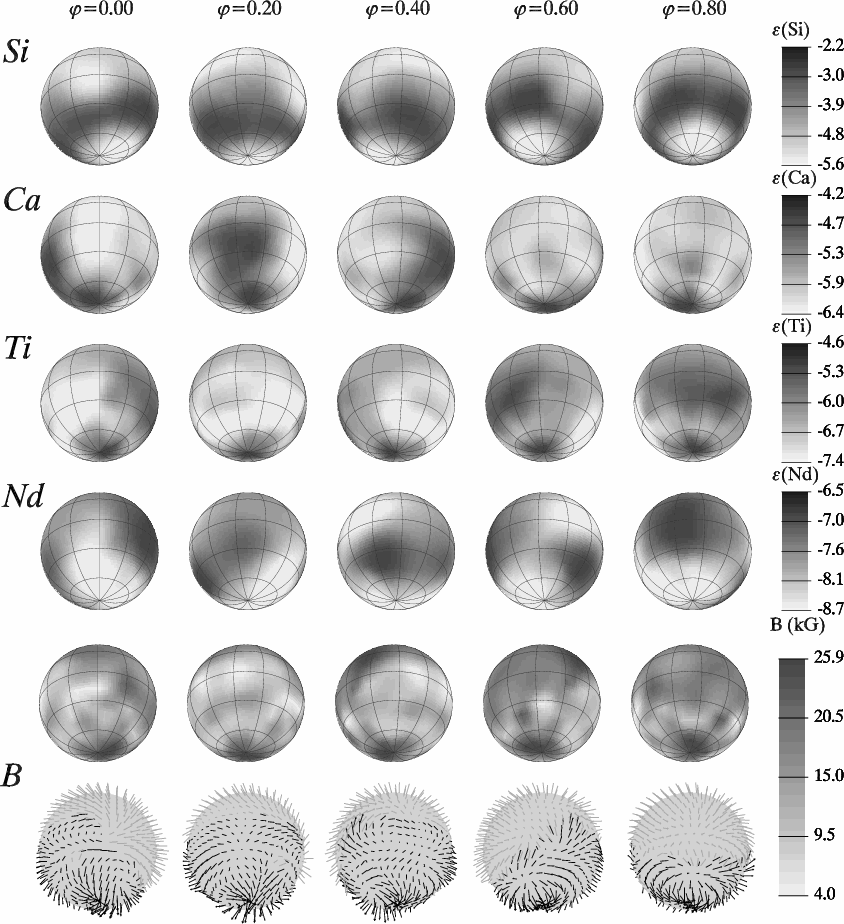}
\vspace{-5mm}
\caption[]{Abundance distributions of Si, Ca, Ti and Nd on the surface of 53 Cam. In the the two lower 
panels the average magnetic map reconstructed from the Fe II lines is shown. (Figure adopted from Kochukhov et al. 2004).}
\label{}
\end{center}
\end{figure}

\begin{figure}[!t]
\begin{center}
 \includegraphics[width=8cm]{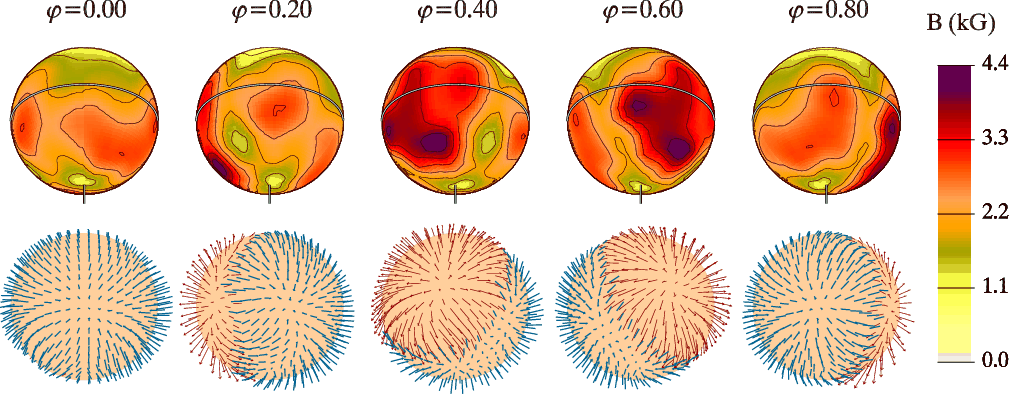}
 \includegraphics[width=8cm]{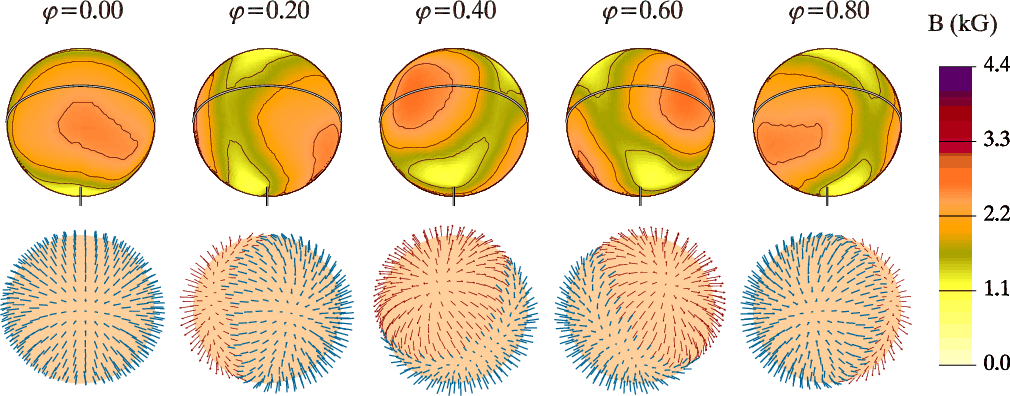}
 \includegraphics[width=8cm]{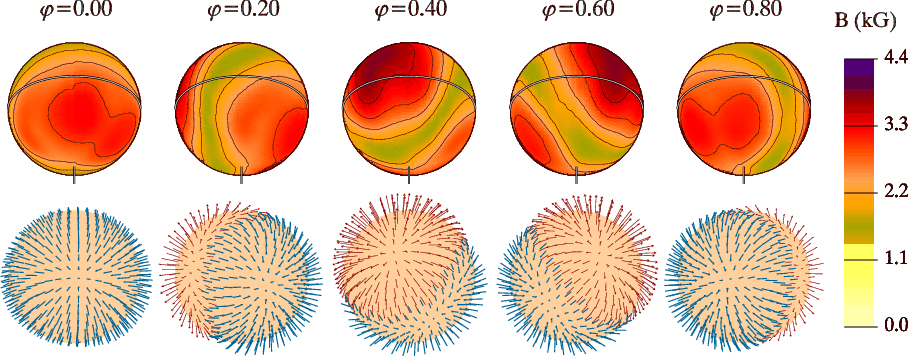}
\vspace{-5mm}
\caption[]{Top panels: Surface magnetic field of $\alpha^2$\,CVn derived from Stokes IQUV (based on Fe II and Cr II lines), 
contours of equal magnetic field strength are plotted every 0.5 kG, 
two panels in the middle: same as above, but 10 times larger Tikhonov regularization, 
two bottom panels: surface magnetic field derived using Stokes I and 
V with multipolar regularization. (Figure adopted from Kochukhov et al. 2010).}
\label{}
\end{center}
\end{figure}

\section{Bayesian photometric Imaging}

Inhomogeneous elemental abundance distributions and the 
resulting spotty structures presumably also cause the stellar photometric variability linked to 
rotation typical for Ap stars. Such kind of structures were recently reported and successfully 
modeled for e.g. HD 37776 (Krti\v{c}ka et al. 2007, Krti\v{c}ka et al. 2012), $\epsilon$\,UMa (Shulyak et al. 2010), and HD 50773 
(Lueftinger et al. 2010a). While brightness variations of solar type stars arise from activity induced by dynamo 
action inside the star and the related temperature spots, the physical nature of photometric variations
in Ap stars is directly connected with the radiative flux redistribution due to enhanced or deficient opacity in 
abundance spots relative to the rest of the stellar surface. Hence, as a star rotates, the 
observer sees different stellar regions that are emitting a different amount of radiative flux, producing 
the characteristic variability of indices in phase-resolved photometry. From the photometric point of 
view, the surface structure of stars to date could only be studied phenomenologically, but with the advent 
of space missions such as CoRoT, MOST, Kepler, and BRITE-C, and applying a Bayesian approach to star spot 
modeling (BPI), we now have the possibility of estimating stellar surface parameters and their uncertainties 
from light curves of unprecedented quality and on excellent time base. A main advantage of a Bayesian approach to 
the photometric modeling of stellar spots is the possibility to estimate 
all necessary parameters and their uncertainties exclusively from observational data without having to rely 
on simulations and artificial datasets. 
Within the Bayesian algorithm, marginal distributions of parameters like inclination angle, stellar rotational 
period, the longitude, latitude, intensity and the radius for each spot need to be determined. 
Once the marginal distribution of a parameter is calculated, all other necessary 
quantities like, e.g., mean, median, mode, standard deviation, and confidence regions follow. For the computationally 
demanding task of assigning mean values and error bars, the Markov-Chain Monte-Carlo (MCMC) technique is used.  

For the CoRoT CP2 target star HD 50773, it was possible (Lueftinger et al. 2010a) to directly investigate the correlation 
of surface brightness patches determined via BPI (based on CoRoT space photometry) 
to the chemical abundance distribution plus magnetic field structure using rotation phase resolved spectropolarimetry and 
MDI. The results presented in Fig. 3 show an astonishingly similar reproduction of the stellar surface from photometry 
(second panel from top) and spectroscopy (panels 3, 4, and 5 from top), and thus from two totally different and independent 
analysis techniques. In Ap star research, we now seem to have a powerful combination of photometric data 
obtained in space with ground based spectroscopy and spectropolarimetry at hands. 

\begin{figure}[!t]
\begin{center}
 \includegraphics[width=8cm]{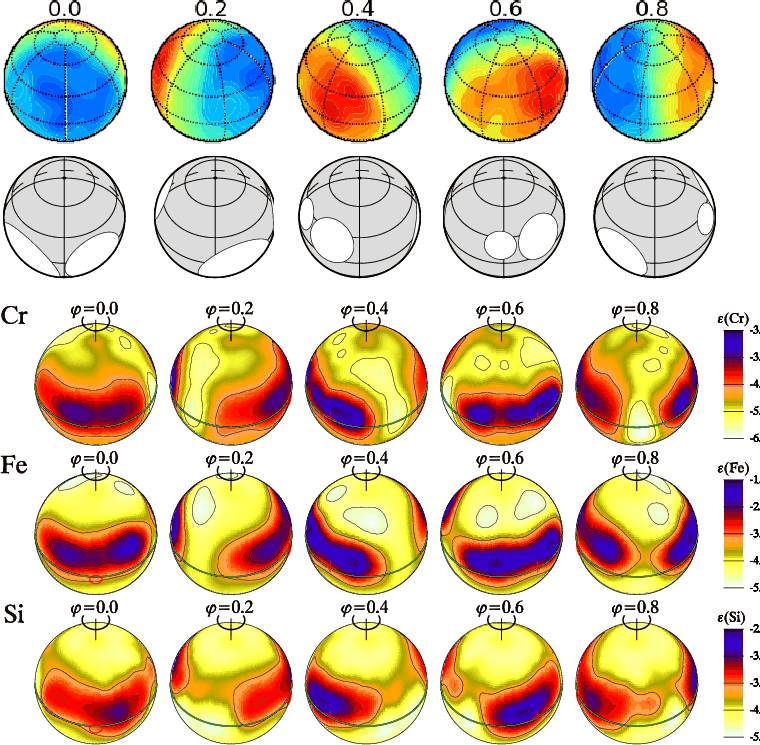}
\vspace{-5mm}
\caption[]{Top panel: radial field component of the magnetic map of HD 50773. 
Second panel: locations of the four bright photometric spots. 
Next three panels: abundance distribution of Cr, Fe, and Si. 
The circle and the cross indicate the position of the negative and 
positive magnetic pole, respectively. 
Applicable to all Figures: Darker areas in the plots correspond to higher elemental abundances, the 
corresponding scale is given to the right of 
each panel. 
All projections are plotted at five equidistant rotation phases. (Figure adopted from Lueftinger et al. 2010).}
\label{}
\end{center}
\end{figure}

\section{Conclusions}

For several decades, DI was applied only to a small number of chemical species, and magnetic fields  
were analysed with coarse techniques, assuming low-order multipolar magnetic topologies inferred from 
longitudinal magnetic field measurement. Recent groundbreaking advances in theoretical modeling, observational instrumentation and
analysis techniques enabled an enormous development of the (M)DI technique, and we now have reached a stage of maturity, where chemical
geometry and magnetic field structures on the surfaces of Ap stars can be determined simultaneously and consistently, and, in the case
of full Stokes parameter inversions, without having to rely on any a priory assumptions. 
These advances have lead to a significant increase of the number of stars mapped. The maps 
reveal that Ap star surfaces show a pronounced diversity of spot structures, still to
be explained. The inclusion of linear Stokes parameters in MDI indicates complex patches of field intensitiy distributions beneath a
dominantly dipolar overall field configuration. The small scale structures due to the horizontal magnetic field component possibly
influence the aforementioned diversity of chemical patches and an apparent lack of correlation to the overall topology for most of the
chemical species.  
Bayesian Photometric imaging since recently offers an additional asset and might help in increasing the statistics of stars analysed
via DI, MDI, and BPI, trying to find an overall picture of the interplay and the relation of diffusion, chemical surface patterns and 
magnetic field geometries in Ap stars' atmopheres. 

\bigskip
{\it Acknowledgements.} This work was supported by the  Austrian Science Fund (FWF-P19962).  
The author thanks O. Kochukhov and T. Ryabchikova for useful discussions. 
Scientific analysis made use orm f the SIMBAD, and VALD databases.

%%%%%%%%%%%
\end{document}